\documentclass[11pt,preprintnumbers,amsmath,amssymb,floatfix]{revtex4}
\usepackage[pdftex]{}
\usepackage{graphicx}
\usepackage{dcolumn}
\usepackage{bm}
 \usepackage {mathrsfs} 
\usepackage{epsfig} 
\usepackage {array} 
\usepackage {amsmath} 
\usepackage {lscape}
\usepackage{color}
\DeclareMathOperator{\Real}{\mathrm{Re}}
\DeclareMathOperator{\Imag}{\mathrm{Im}}

\newcommand{\Rm}{\mathrm{Rm}}

\providecommand\bB{\bm{{\rm B}}}

\providecommand\br{b_{r}}
\providecommand\bphi{b_{\theta}}

\newcommand{\Ds}{\partial_{u}}

\newcommand{\Dt}{\partial_t}
\newcommand{\Dr}{\partial_r}

\renewcommand{\L}{\mathscr{L}}
\newcommand{\Dtob}{\partial_{\bar{\tau}}}
\newcommand{\Dto}{\partial_{\tau}}

\newcommand{\brho}{\hat{b}_{r0}}
\newcommand{\bphiho}{\hat{b}_{\theta0}}

\begin{document}
\preprint{}
\title{Oscillating Ponomarenko dynamo in the highly conducting limit}
\author{Marine Peyrot$^{1}$, Andrew Gilbert$^{2}$, Franck Plunian$^{1}$}
\email{Marine.Peyrot@ujf-grenoble.fr, Franck.Plunian@ujf-grenoble.fr, A.D.Gilbert@ex.ac.uk}
\affiliation{$^{1}$ Laboratoire de G\'eophysique Interne et Tectonophysique, Universit\'e Joseph Fourier, CNRS, Maison des G\'eosciences, B.P. 53, 38041 Grenoble Cedex 9, France\\
$^{2}$ Mathematics Research Institute, School of Engineering, Computing and Mathematics,\\University of Exeter, Exeter, EX4 4QF, U.K.}
\date{\today}

\begin{abstract}
This paper considers dynamo action in smooth helical flows in cylindrical geometry, otherwise known as Ponomarenko dynamos, with periodic time dependence. An asymptotic framework is developed that gives growth rates and frequencies in the highly conducting limit of large magnetic Reynolds number, when modes tend to be localized on resonant stream surfaces. This theory is validated by means of numerical simulations.
\end{abstract}

\maketitle

\section{Introduction}

A well-known kinematic dynamo model goes back to the work of Ponomarenko \cite{Ponomarenko73}, who found that magnetic modes could be amplified in a flow field in cylindrical geometry (depending only on distance from the axis), which generally possesses helical streamlines. In recent studies this Ponomarenko dynamo has been investigated when the helical flow is modulated in time \cite{Normand03, Peyrot07}, with a focus on the dynamo threshold. The aim here is to quantify the effect of simple time-periodic fluctuations on the mean flow, and the effect of these on the threshold for magnetic growth. The main conclusion is that the dynamo threshold is larger than the one obtained without fluctuations, suggesting that large scale fluctuations are not desirable when optimizing a dynamo experiment. In dynamo experiments, such large scale fluctuations have been avoided simply by adding inner walls \cite{Gailitis01, Stieglitz01} or a flow-stabilizing ring \cite{Monchaux06}. A further experiment in preparation \cite{Frick02} is based on a single helical flow, again avoiding large scale fluctuations. 

The above simulations \cite{Normand03,Peyrot07} were purely numerical and in order to give some theoretical backing to the results it is necessary to use an asymptotic limit where approximate results can be obtained. The steady Ponomarenko dynamo \cite{Ponomarenko73} has been studied for $\Rm \gg 1$ in the kinematic case \cite{Gilbert88, Ruzmaikin88, Gilbert00} and equilibrated solutions have been found taking into account the nonlinear feedback on the flow \cite{Dobler02}. In both cases the underlying flow is steady, and our aim here is to extend the kinematic results to the case of oscillatory flow fields. 

We therefore adopt the limit of large magnetic Reynolds number ($\Rm\gg 1$), generally much above the threshold. Although our aim is to derive asymptotic results which are valid for this general class of flows, with an eye to experimental dynamos and numerical simulations, we note that Ponomarenko dynamos may also occur in bipolar jet-like outflows commonly observed in protostellar systems, in which magnetic field probably plays an important role. Though it is argued that such magnetic fields are produced inside the protostellar disk \cite{Blackman04} we cannot exclude the existence of a Ponomarenko type dynamo in such a helical jet in which strong time fluctuations may also occur. 

Here we assume a time-periodic flow depending only on radius, limit our investigation to the kinematic approximation, and study the asymptotic limit of large $\Rm$ (section \ref{sec2}). Our analysis will be compared to direct numerical simulations for three cases (section \ref{sec3}): stationary flow and oscillatory flow with zero mean flow (ZM), or non-zero mean flow (NZM). Finally in section \ref{sec4} we generalize our analysis to the case of time-varying resonant radius.

\section{Model and asymptotic approximation}
\label{sec2}

The time evolution of the magnetic field is given by the dimensionless induction equation
\begin{equation}
	\varepsilon  \frac{\partial \textbf{B}}{\partial t} = \nabla \times (\textbf{U} \times \textbf{B}) + \varepsilon  \nabla^2 \textbf{B}, 
\label{induc}
\end{equation}
with $\varepsilon = \Rm^{-1}$ and where we have adopted a diffusion time scale for our time variable (in contrast to \cite{Gilbert88}, but in accord with \cite{Peyrot07}). We consider a time-dependent helical flow expressed in cylindrical coordinates $(r, \theta, z)$ by 
\begin{equation}
 \textbf{U}(r,t)=(0,r\Omega(r),V(r)) F(t) \quad \mbox{for} \quad r\le 1, \quad \quad \quad
 \textbf{U} ={0} \quad \mbox{for} \quad r > 1 \label{velocity} , 
 \end{equation}
where $\Omega$ and $V$ are smooth functions of $r$ and $F$ is a given function of time. In the stationary case $F=1$, and for the non-stationary case we will consider two functional forms for the time-dependence,
\begin{equation}
F(t) = \cos\omega t \quad \text{(ZM)}, \qquad
F(t) = 1 + \rho\cos\omega t \quad \text{(NZM)}, 
\label{eqFforms}
\end{equation}
the `zero-mean' and `non-zero mean' flows, respectively. 

For the linear, kinematic dynamo problem we may consider a magnetic field of the form 
  \begin{equation}
  \textbf{B}(\textbf{r},t)= e^{i(m \theta + k z + \phi(t))} \,   \textbf{b}(r,t) , 
  \label{B}
  \end{equation}
where $m$ and $k$ are the azimuthal and vertical wave numbers of the field and $\phi(t)$ is a phase, put in for convenience, that we will choose shortly. The solenoidality of the field, $\nabla \cdot \textbf{B} = 0$, expressed in cylindrical coordinates, 
\begin{equation}
b_r' + r^{-1}{b_r} +  {im}{r}^{-1}b_{\theta} + ik b_z = 0 , 
\end{equation}
shows that it is enough to solve the induction equation for ${b_r}$ and $b_{\theta}$ only.

It has been shown \cite{Gilbert88,Ruzmaikin88} that for a stationary flow the magnetic field is generated in a resonant layer located at $r=r_0$, where the magnetic field lines are aligned with the shear and thus minimize their diffusion. 
In equation (\ref{velocity}) we see that since there is the same time-dependent factor $F(t)$ multiplying angular and axial velocities, this radius is independent of time, and so this surface is fixed and given by 
\begin{equation}
	m\Omega'(r_0) + k V'(r_0) = 0.
	\label{resonnance}
\end{equation}
For more complex time-dependence $r_0$ may vary with time, leading to a succession of growing and damping magnetic field states \cite{Peyrot07}. In section \ref{secres} of this paper we will consider the velocity field in (\ref{velocity}) with $r_0$ fixed, and we assume that $r_0$ lies in the fluid, with $0<r_0<1$ (otherwise modes are strongly damped). For the more general case of distinct time-dependence for angular and axial flows, where $r_0$ does vary with time, the equations are set out and discussed in section \ref{secnonres}. 

With a given resonant surface $r=r_0$ the leading action of the flow on the field is simply advection of the mode by the angular and axial velocities, on the fast advective timescale $t=O(\varepsilon)$. We take this out of consideration by defining the phase $\phi(t)$ as
\begin{equation}
\phi(t)=- \varepsilon^{-1} \left( m\Omega(r_0) + k V(r_0) \right) \int\limits_{0}^{t}F(t')dt' , 
\label{phi}
\end{equation}
to leave behind only evolution through dynamo action, diffusion and reconnection, on slower time-scales. 

Introducing (\ref{velocity}), (\ref{B}) and (\ref{phi}) in (\ref{induc}), we find
\begin{eqnarray}
\lbrack \varepsilon \Dt &+& \left( i  m  \Omega(r)  +  i k  V(r)- i   m  \Omega(r_0)  -  i k  V(r_0) \right) F(t)  \rbrack  \br  \nonumber \\
&=&  \varepsilon \left( \left(    \L-  r^{-2} \right)  \br  -2  i  m   r^{-2}  \bphi  \right)  , \label{2rbis}  \\
\lbrack \varepsilon \Dt &+& \left( i  m  \Omega(r)  + i k  V(r)-im\Omega(r_0)  -  i k  V(r_0) \right) F(t)   \rbrack  \bphi  \nonumber \\
&=& r   \Omega'(r)   F(t)  \br  + \varepsilon \left( \left(    \L-  r^{-2} \right)  \bphi  +2  i  m   r^{-2}  \br  \right) , 
\label{2phibis}  
\end{eqnarray}
where the Laplacian operator $\L$ is defined by
\begin{equation}
\label{op}
\L= \Dr^{2}+r^{-1} \Dr  - {r^{-2}}{m^{2}} - k^{2} .
\end{equation}
In the highly conducting limit $\varepsilon \ll 1$ we adopt the smooth Ponomarenko dynamo scaling \cite{Gilbert88},
\begin{equation}
m=\varepsilon^{-1/3}M,   \quad  k=\varepsilon^{-1/3}K ,    \quad  
r=r_0+\varepsilon^{1/3}s ,  \quad t=\varepsilon^{2/3}\tau  , 
\label{scale}
\end{equation}
where $\tau$ is a time-scale on which the dynamo mode grows, intermediate between the $O(1)$ diffusive time-scale and the $O(\varepsilon)$ advective time-scale. This scaling gives a magnetic mode localized at the radius $r=r_0$ and it is known that the final formulae obtained with this choice of scaling give the `richest' asymptotic picture including both the case $m,k=O(1)$ and the peak growth rates, achieved at $m,k=O(\varepsilon^{-1/3})$. Setting
\begin{equation}
\br (r,t) = \varepsilon^{1/3}  \, \brho (s,\tau) + \cdots, \quad 
\bphi (r,t) =  \bphiho (s,\tau) + \cdots , \quad 
F(t)= \hat{F}(\tau)  , 
\label{Bscale}
\end{equation}
together with the $\Omega(r)$ and $V(r)$ expansion at $r=r_0$,
\begin{eqnarray}
\Omega(r) &=&\Omega(r_0) + \varepsilon^{1/3}  s  \Omega'(r_0)+ \tfrac{1}{2}    \varepsilon^{2/3} s^{2}  \Omega''(r_0)  + \dots ,  \label{expandO}\\
V(r) &=&V(r_0) + \varepsilon^{1/3}  s  V'(r_0)+ \tfrac{1}{2}    \varepsilon^{2/3} s^{2}  V''(r_0)  + \dots , 
\label{expandV}
\end{eqnarray}
we obtain from (\ref{2rbis}) and (\ref{2phibis}) at leading order in $\varepsilon$, 
\begin{eqnarray}
\lbrack \Dto + c_{0} + i c_{2} s^{2} \hat{F}(\tau) -  \partial_s^{2}  \rbrack    \brho &=& {-2iM}{r_0^{-2}} \bphiho , 
  \label{3r} \\
\lbrack \Dto + c_{0} + i c_{2} s^{2} \hat{F}(\tau) -  \partial_s^{2}  \rbrack      \bphiho &=& r_0 \Omega'(r_0)  \hat{F}(\tau) \brho , 
  \label{3phi} 
\end{eqnarray}
with
\begin{equation}
c_{0}= {r_0^{-2}}{M^{2}}  + K^{2}, \quad \quad \quad
c_{2}= \tfrac{1}{2} \left( M\Omega''(r_0)+K  V''(r_0) \right) .  \label{c}
\end{equation}
From (\ref{3r}) and (\ref{3phi}) we immediately see how the dynamo works: the differential rotation $\Omega'(r_0)$ stretches radial field $\brho$ to generate $\bphiho$, and the diffusion of $\bphiho$ in cylindrical geometry then regenerates $\brho$. For the flows considered below $c_2>0$, and for simplicity we will take this to be the case in what follows: there are insignificant changes if this quantity is negative. 

These equations are solved by an exact ansatz involving time-dependent, complex Gaussian functions. We put these in, at the same time rescaling using constants $a_r$, $a_\theta$, $a_\tau$ and $a_h$, to eliminate as many parameters as possible, with 
\begin{equation}
	\left( \begin{array}{l} \brho(\tau,s)\\ \bphiho (\tau,s) \end{array} \right) 
	\exp(  c_0 {\tau}) = 
	\left(  \begin{array}{l} a_r \bar{f}(\bar{\tau}) \\  
                                              a_\theta \bar{g}(\bar{\tau}) 
                    \end{array} \right)
	\exp( -  a_h \bar{h}(\bar{\tau})s^2) , 
	\label{layer}
\end{equation}
where $\bar{\tau}=a_\tau \tau$ is a new time scale.  We now fix the constants 
with
\begin{equation}
a_r  = 2M , \quad
a_\theta = r_0^2 c_2^{1/2} , \quad
a_\tau = a_h = c_2^{1/2}, 
\end{equation}
and we are left with the following system of ODEs in $\bar{\tau}$ to solve
\begin{eqnarray}
\Dtob \bar{h}  +  4  \bar{h}^{2} &=& i   \bar{F}(\bar{\tau})  , \nonumber \\
\Dtob  \bar{f} +  2  \bar{h}  \bar{f} &=& -  i  \bar{g} , \label{f3} \\
\Dtob  \bar{g} +  2  \bar{h}  \bar{g} &=& -  \mathscr{D}  \bar{F}(\bar{\tau}) \bar{f} , \nonumber
\end{eqnarray}
where the constant $\mathscr{D}={-2  M  \Omega'(r_0)}/ r_0 c_{2}$ and  the time-dependent factor is $\bar{F}(\bar{\tau})=\hat{F}(\tau)$.  (Note that modes with more radial structure can be studied, taking the form of two time-dependent polynomials times a Gaussian in (\ref{layer}), but these will be subdominant).

The equation for $\bar{h}$ is nonlinear, while those for $\bar{f}$ and $\bar{g}$ are linear. The exponential growth rate $\bar{\gamma}$ of the magnetic field components $\bar{f}$ and $\bar{g}$ depends only on the parameter $\mathscr{D}$, which captures the local geometry of the helical streamlines  \cite{Gilbert88}, and the form of the time-dependence $\bar{F}(\bar{\tau})$. From (\ref{resonnance}) and (\ref{c}) we have
\begin{eqnarray}
\mathscr{D}=- \frac{4}{r_0}\left(\frac{\Omega''(r_0)}{\Omega'(r_0)}- \frac{V''(r_0)}{V'(r_0)} \right)^{-1}  , 
\label{scriptD}
\end{eqnarray}
showing that $\mathscr{D}$ depends only on the geometry of the velocity field.
The function $\bar{h}(\bar{\tau})$ gives the Gaussian envelope, with $\Real \bar{h}>0$ required for exponential localization of the mode. 

So far the system (\ref{f3}) applies to any time-dependence $\bar{F}(\bar{\tau})$.
Under the rescaling, the two given in (\ref{eqFforms}) correspond to 
\begin{equation}
\bar{F}(\bar{\tau}) = \cos\bar{\omega}\bar{\tau} \quad \text{(ZM)}, \qquad
\bar{F}(\bar{\tau}) = 1 + \rho\cos\bar{\omega}\bar{\tau}  \quad \text{(NZM)}, 
\label{eqforcrescale}
\end{equation}
where the frequencies are linked by 
\begin{equation}
\omega = \varepsilon^{-2/3} c_2^{1/2}  \, \bar{\omega} 
 \equiv  [\tfrac{1}{2}\varepsilon^{-1} (m\Omega''(r_0) + kV''(r_0))]^{1/2} \, 
        \bar{\omega} . 
        \label{linkfreq}
\end{equation}

For the rescaled system (\ref{f3}), after a transient, $\bar{h}$ will become periodic with the same frequency $\bar{\omega}$ as the forcing $\bar{F}$ and the linear equations for the magnetic field components $(\bar{f}, \bar{g})$ will take a Floquet form: the solution will have exponential growth superposed on periodic behaviour. The overall growth rate may be measured as $\bar{\gamma}(\mathscr{D},\bar{\omega})$ (suppressing $\rho$ in the non-zero mean case). This is linked to the growth rate $\gamma$ of magnetic field in the original system (\ref{2rbis}), (\ref{2phibis}) (or (\ref{induc})) with
\begin{equation}
\gamma = \varepsilon^{-2/3} [c_2^{1/2}  \bar{\gamma}(\mathscr{D},\bar{\omega}) 
- c_0 ]  
\equiv  [ \tfrac{1}{2} \varepsilon^{-1} ( m  \Omega''(r_0)  +  k  V''(r_0)) ]^{1/2}  \, 
\bar{\gamma}(\mathscr{D},\bar{\omega}) -  r_0^{-2} m^{2 } -  k^{2} , 
\label{growthrate}
\end{equation}
or, using the definition of $\mathscr{D}$ in (\ref{scriptD}),
\begin{equation}
\gamma =  [ -2 m  \Omega'(r_0)/\varepsilon r_0\mathscr{D} ]^{1/2}  \, 
\bar{\gamma}(\mathscr{D},\bar{\omega}) -  r_0^{-2} m^{2 } -  k^{2} . 
\label{growthrateD}
\end{equation}

In the stationary case $F(t)=1$, from (\ref{f3}) we find  $\bar{h}=\pm i^{1/2}/2$ and $\bar{\gamma}=-2\bar{h} \pm ({i\mathscr{D}})^{1/2} $.  Then from (\ref{growthrate}) and taking $\bar h$  with $\Real\bar h\ge 0$, we obtain the real part of the growth rate as 
\begin{equation}
\Real \gamma = \varepsilon^{-1/2}   (r_0^{-1} | m \Omega'(r_0)|)^{1/2}  -\tfrac{1}{2} \varepsilon^{-1/2}
( |m  \Omega''(r_0)+ k V''(r_0) |)^{1/2}  -  r_0^{-2}{m^{2}}  - k^{2}  , 
\label{growthrateagain} 
\end{equation}
as previously found \cite{Gilbert88,Ruzmaikin88, Gilbert00}. Together with this goes the purely geometrical criterion for Ponomarenko dynamo action in highly conducting stationary flow, that $|\mathscr{D}(r_0)| > 1$ at the resonant radius $r_0$. Note that  formulae (\ref{linkfreq})--(\ref{growthrateagain}), although derived using the scaling (\ref{scale}), are in fact valid for all $m$ and $k$ linked by the resonance condition (\ref{resonnance}). The expansions would give equivalent results had we taken $m$, $k=O(1)$ and $\varepsilon\to0$, though this would not immediately capture the fastest growing modes which are of the scale $m$, $k=O(\varepsilon^{-1/3})$ as $\varepsilon\to0$. The key assumption in the expansion is that at small $\varepsilon$ the magnetic field localises in a thin layer. From (\ref{scale}) and (\ref{layer}) the width of the layer is 
\begin{equation}
 r-r_0 = O\left(\varepsilon^{1/3} c_2^{-1/4} \bar{h}^{-1/2} \right) 
  = O\left(\varepsilon^{1/4} [\tfrac{1}{2} (m\Omega''(r_0 + kV''(r_0))]^{-1/4} \bar{h}^{-1/2} \right) . 
\label{layerwidth} 
\end{equation}
and this goes to zero as $\varepsilon\to0$; however we should note that this is assuming that $\bar{\omega}$ is fixed as we take the limit, so that the magnitude of $\bar{h}$ is of order unity in the limit. If instead we allow $\bar{\omega}$ (or other parameters or wavenumbers) to vary as well, then we need to be careful to check the condition that the width given by (\ref{layerwidth}) is small, to validate the asymptotic theory. For example if we fix $\omega$ as $\varepsilon\to0$ we have $\bar{\omega} = O(\varepsilon^{1/2}\omega)\to0$ from (\ref{linkfreq}) and as $\bar{h}$ turns out to be bounded in this limit (of low frequencies, so similar to the stationary case) this condition is verified. 

\section{Results}\label{secres}
\label{sec3}
To test the above asymptotic results, we use the flow (\ref{velocity}) with radial profile $\Omega(r)=1-r$ and $V(r)=\Gamma(1-r^2)$ where $\Gamma$ is a helicity factor, for which $r_0=-m/2k\Gamma$ and $\mathscr{D}=4$, independent of radius. We begin by checking the stationary case, followed by the examples of zero-mean and non-zero mean flows in (\ref{eqFforms}). 

The growth rate of magnetic field for the asymptotic theory is obtained by simulating (\ref{f3}) using a fourth order Runge--Kutta scheme, with frequencies and growth rates linked by (\ref{linkfreq}) and (\ref{growthrate}). For our given flow we have 
\begin{equation}
	\omega = (-k\Gamma \varepsilon^{-1})^{1/2} \bar{\omega}, \quad \quad 
	\gamma = (-k\Gamma \varepsilon^{-1})^{1/2} \bar{\gamma} -(1+4\Gamma^2)k^2 . 
	\label{omgamma}
\end{equation}
Typically for $\Gamma=2$ and $k=-0.5$, we have $\omega=\varepsilon^{-1/2}\bar{\omega}$ and $\gamma = \varepsilon^{-1/2}\bar{\gamma}-4.25$.
This growth rate will be compared to the one obtained with direct numerical simulation, solving the induction equation (\ref{induc}) without asymptotic approximation, using a Galerkin method for the radial discretization and again a Runge--Kutta scheme for the time evolution \cite{Peyrot07}. Note that for the full problem the field settles into a Floquet form with $\bB(t+T)=\exp(\gamma_BT)\bB(t)$, 
where $T$ denotes the period of $F(t)$.  From (\ref{B}) and (\ref{phi}) it is given by
\begin{equation}
\gamma_B = \gamma -i(m\Omega(r_0))+kV(r_0))\, {T}^{-1} \int^{T}_{0}F(t')\, dt'. 
\end{equation}
where the phase factor has been reintroduced for a good comparison of frequency measurements. 

\subsection{Stationary flow}
\begin{figure}
   \includegraphics[width=1\textwidth]{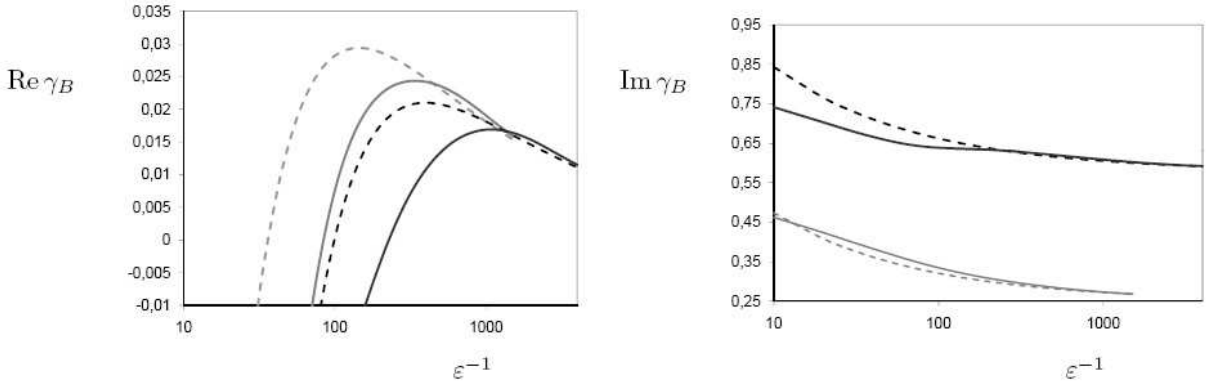}
\caption{The magnetic growth rate $\Real\gamma_B$ (left) and frequency $\Imag \gamma_B $ (right) plotted against $\varepsilon^{-1}$ in the stationary case $F(t)=1$, for $m=1$, $\Gamma = 2$ and $k=-0.7$, $r_0=0.35$  (black), and $k=-0.5$,  $r_0=0.5$ (grey). The asymptotic solution and the simulation correspond to dashed and full curves respectively.}  
\label{fig:stat}
\end{figure}
\begin{figure}
    \includegraphics[width=1\textwidth]{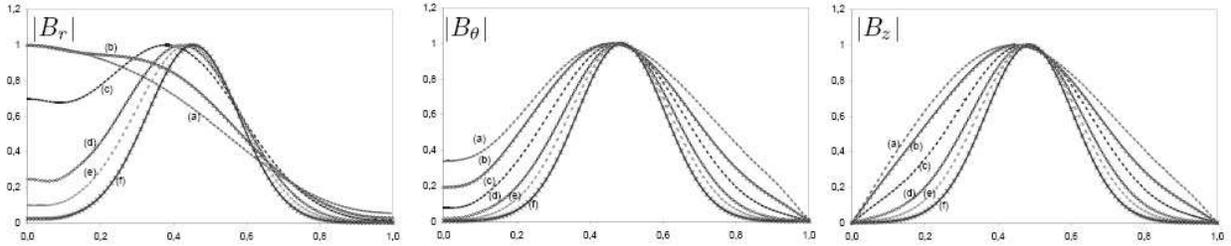}
\caption{ Modulus of each magnetic field component plotted against $r$, for (a) $\varepsilon^{-1}=  500$,  (b) 1000, (c) 2000, (d) 4000, (e) 6000, (f) 10000, for $m=1$, $\Gamma = 2$, $k=-0.5$ and $r_0=0.5$.}  
\label{fig:B_fc_r}
\end{figure}
For $F(t)=1$ and our given flow, we have $\bar{\gamma}=(1+i)/\sqrt{2}$. From  (\ref{omgamma}) we obtain $\gamma_B = \gamma -i(m\Omega(r_0))+kV(r_0))$ with
\begin{equation}
\gamma=\varepsilon^{-1/2} \sqrt{\tfrac{1}{2} |k|\Gamma} \, \,(1+i) -  \left(1+ 4\Gamma^2 \right)k^{2}. 
\label{growthrate(i)}
\end{equation}
In figure \ref{fig:stat}, the growth rate $\gamma_B$ is plotted against $\varepsilon^{-1}$ for two different values of $r_0$. The curves show a good agreement between both asymptotic and simulation growth rates and frequencies provided the magnetic Reynolds number $\Rm = \varepsilon^{-1}$ is sufficiently large. For $\varepsilon^{-1} \ge 10^3$ the difference is less than $5 \%$ for the growth rate and $0.6 \%$ for the frequency. In figure \ref{fig:B_fc_r}, the modulus of each magnetic field component is plotted for several values of $\varepsilon^{-1}$. Clearly, increasing $\varepsilon^{-1}$ concentrates the magnetic field in a  thinner layer at $r_0$, and the asymptotic formulation becomes increasingly accurate. Defining the layer thickness $\delta$ as the width over which the magnetic energy falls to half of its peak value, we find that $\delta \simeq O(\varepsilon^{0.27 \pm 0.03})$ from the simulation. An estimate from the asymptotic expressions (\ref{layer}), and using (\ref{scale}) and (\ref{c}), leads to $\delta =O( \varepsilon^{1/4})$ for $m=1$ which is in good agreement. However this quantity goes to zero quite slowly with $\varepsilon$ and so convergence is slow. 


\subsection{Periodic flow with zero mean}

\begin{figure}
    \includegraphics[width=1\textwidth]{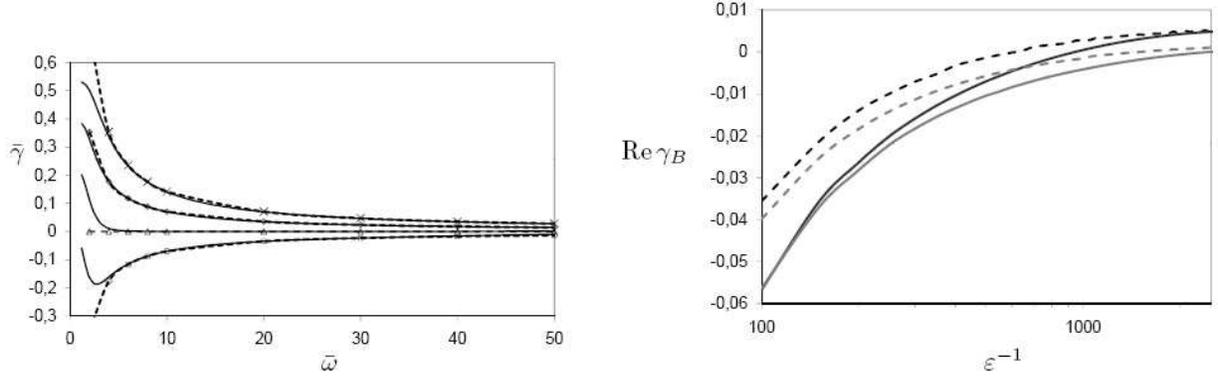}
\caption{Zero mean case: growth rate $\bar{\gamma}$ versus $\bar{\omega}$  (left) for, from bottom to top, $\mathscr{D}=1, 2, 3, 4$. The full curves give growth rates from integration of (\ref{f3}); dashed curves and symbols give growth rates from (\ref{growthhighfreq}).
Growth rate $\Real \gamma_B$ versus $\varepsilon^{-1}$ (right) for $m=1$, $\Gamma = 2$, $k=-0.5$,  $r_0=0.5$ and $\omega=200$ (black) $\omega=500$ (grey).
The asymptotic results are shown by dashed curves, simulations by full curves.}
\label{fig:ZeromeanODE}
\end{figure}

We now consider the second case, with zero mean in the original time-dependence (\ref{eqFforms}) or in the rescaled version (\ref{eqforcrescale}). We first work with the asymptotic system and solve (\ref{f3}) to obtain the growth rate $\bar{\gamma}(\mathscr{D},\bar{\omega})$ as a function of $\bar{\omega}$ for different values of $\mathscr{D}$, as plotted in figure \ref{fig:ZeromeanODE} (left).
For $\mathscr{D}<1$ we obtain pure decay $\Real\bar{\gamma}<0$; for $\mathscr{D}= 1.5$ the sign of $\Real \bar{\gamma}$ depends on $\bar{\omega}$ (positive at small $\bar{\omega}$) whereas for $\mathscr{D}\ge 2$ we find that $\Real\bar{\gamma} \ge 0$ for all $\bar{\omega}$. Recall that in the stationary case $|\mathscr{D}|>1$ is necessary and sufficient for dynamo action in the highly conducting limit. 

We can investigate this result further by taking an additional limit of solving the equations (\ref{f3}) for $\bar{\omega}\gg 1$. For that we use a new time coordinate $u = \bar{\omega}\bar{\tau}$ and set a small parameter $\zeta=\bar{\omega}^{-1}\ll1 $. Then we have, without approximation, from (\ref{f3}) and dropping the bars to ease notation, 
\begin{eqnarray}
\zeta^{-1} \Ds  h + 4 h^{2}&=&i\cos u,  \label{equation1h}\\
\zeta^{-1} \Ds  f + 2 h f&=&-ig, \label{equation1f}\\
\zeta^{-1} \Ds  g + 2 hg&=&-\mathscr{D}f \cos u, \label{equation1g}
\end{eqnarray}
where $f$, $g$ and $h$ are now functions of $u$. We set
$(f(u),g(u))=\exp(\mu u)(f^{*}(u), g^{*}(u))$ with $\mu$ a constant Floquet exponent and require $f^{*}$ and $g^{*}$ to be strictly periodic functions of $u$. We have
\begin{eqnarray}
\zeta^{-1} \Ds  h + 4 h^{2}&=&i\cos u , \nonumber \\
\zeta^{-1} \mu  f^{*} +\zeta^{-1}  \Ds  f^{*}+2 h f^{*}&=&-ig^{*} , \label{equation2f}\\
\zeta^{-1} \mu  g^{*} +\zeta^{-1}  \Ds  g^{*}+2 h g^{*}&=&-\mathscr{D}f^{*}\cos u . \nonumber 
\end{eqnarray}
Expanding $\mu$, $f^{*}$, $g^{*}$ and $h$ in powers of $\zeta$, 
\begin{equation}
	(\mu, f^{*}, g^{*}, h)=(\mu, f^{*}, g^{*}, h)_0 + \zeta (\mu, f^{*}, g^{*}, h)_1 + \zeta^2 (\mu, f^{*}, g^{*}, h)_2 + \cdots, 
\end{equation}
we solve the system (\ref{equation2f}) order by order, using the terms $\mu_0$, $\mu_1, \ldots$ to enforce periodicity. This leads to
\begin{eqnarray}
f^{*}&=&A_{0}+\zeta A_{1}+\zeta^{2}(iA_{0}(2-\mathscr{D})\cos u+A_{2})+\cdots,
\label{fseries}\\
g^{*}&=&\zeta\mathscr{D}A_{0}(-\sin u\pm 2^{-1/2}i)+\zeta^{2}(-\mathscr{D}A_{1}\sin u+B_{2})+\cdots , 
\label{gseries}\\
h&=&\zeta(2^{-1/2}+i\sin u)+\zeta^2 C_2 + \zeta^{3}(C_{3}-\sin 2u+4\sqrt{2}\, i \cos u)+\cdots , 
\label{hseries} \\
\mu&=&\zeta^{2}(-{2}^{1/2}\pm2^{-1/2}{\mathscr{D}})+\cdots , 
\end{eqnarray}
where the $A_i$, $B_i$ and $C_i$ are integration constants and we have imposed $\Real h >0$. (The $A_i$ are arbitrary; the $B_i$ and $C_i$ can be determined in terms of the $A_i$ at higher orders of the expansion.)
Then the growth rate of $\bar{f}$ and $\bar{g}$ is given by $\bar{\gamma}=\mu \bar{\omega}$ with 
\begin{equation}
\bar{\gamma}\simeq\bar{\omega}^{-1} 2^{-1/2} (-2 \pm \mathscr{D}) . 
\label{growthhighfreq}
\end{equation}
This confirms that $\bar{\gamma} \ge 0$ only if $\mathscr{D}\ge 2$, in the high frequency limit, as seen in figure \ref{fig:ZeromeanODE} (left). In addition to the results of system (\ref{f3}), $\bar{\gamma}$ is plotted versus $\bar{\omega}$ using the  asymptotic expansion (\ref{growthhighfreq}). We find a good agreement between both, even for moderate values of $\bar{\omega}$. 
In figure \ref{fig:ZeromeanODE} (right), we plot the growth rate $\gamma_B$ for different values of $\omega$, from both the asymptotic ODEs (\ref{f3}) and from simulations of the primitive equations (\ref{2rbis}), (\ref{2phibis}), showing good agreement provided $\varepsilon^{-1}$ is sufficiently large.

Note that formula (\ref{growthhighfreq}) indicates a growth rate $\bar{\gamma}$ that increases as the frequency $\bar{\omega}\to0$, and this perhaps suggests fast dynamo action. The dynamo here would be fast if the growth rate on the short, advective time-scale, here given by $\varepsilon\gamma$, remains bounded above zero as $\varepsilon\to0$, holding the flow fixed. In our flow  $\mathscr{D}=4$ which, from (\ref{growthhighfreq}), leads to $\bar{\gamma}\sim\sqrt{2}/\bar{\omega}$ and from (\ref{omgamma}) to 
\begin{equation}
\gamma \simeq -\sqrt{2}k\Gamma\varepsilon^{-1}\omega^{-1} - (1+4\Gamma^2)k^2 .
\label{gamma}
\end{equation}
In the limit of small $\varepsilon$, given that $-k\Gamma>0$, this at first sight appears to be a fast dynamo. This formula was derived on the assumption that $\bar{\omega}^{-1}\ll 1$, but as $\varepsilon \to0$ for a fixed flow, which includes a fixed $\omega$, the assumption becomes violated. As $\varepsilon\to0$, $\bar{\omega}
\to0$ from (\ref{omgamma}) and so we move towards the left on figure 3 (left panel): if the asymptotic curves (dashed) continued to grow to the left, the dynamo would be fast.
However the computed values (solid) saturate for small $\bar{\omega}$ and the dynamo is in the slow camp, as expected. (As line elements are only stretched linearly the flow fails to have Lagrangian chaos, technically positive topological entropy, a requirement for fast dynamo action in a smooth flow \cite{Klapper95}.)
 

\subsection{Critical values for $\Rm$ in flows with zero mean}

The asymptotic theory gives an estimate for the critical value of $\Rm$ or $\varepsilon$ for the onset of dynamo instabilities, namely from (\ref{growthrate}), 
\begin{equation}
\varepsilon_{\rm c}^{1/2}  =  \Real \{ [ \tfrac{1}{2}  ( m  \Omega''(r_0)  +  k  V''(r_0)) ]^{1/2} \, 
\bar{\gamma}(\mathscr{D},\bar{\omega}) \}  \,( r_0^{-2} m^{2 } +  k^{2})^{-1} , 
\label{critical1}
\end{equation}
or from (\ref{growthrateD}),
\begin{equation}
\varepsilon_{\rm c}^{1/2}  =  \Real \{ [ -2m\Omega'(r_0) /r_0 \mathscr{D} ]^{1/2} \, 
\bar{\gamma}(\mathscr{D},\bar{\omega}) \}  \,( r_0^{-2} m^{2 } +  k^{2})^{-1}.
\label{critical1b}
\end{equation}

For the mode $m=1$, the agreement with numerics at onset is poor in figure \ref{fig:stat} because the critical magnetic Reynolds number is not large enough (contrast the situation in \cite{Gilbert00}). In the zero-mean case the agreement seems to be better as seen in figure \ref{fig:ZeromeanODE} (right). In fact we generally expect agreement for critical values to improve if there is some other asymptotic parameter to push the critical $\Rm$ into the small-$\varepsilon$, large-$\Rm$ regime, provided the condition of thin layer width (\ref{layerwidth}) is satisfied. One possibility could be to take the the limit when the axial flow $V(r)$ is weak or strong compared with the angular velocity $\Omega(r)$ (as measured by the helicity factor $\Gamma$). Unfortunately direct simulations for $\Gamma \ll 1$ or $\Gamma\gg 1$  are difficult to achieve. Instead we consider the limit of increasing frequency $\omega$. We have already seen that the asymptotic analysis leads to (\ref{gamma}) provided $\bar{\omega}$ is large enough. In this case, the threshold should scale as $\varepsilon_c^{-1}\propto\omega$ and $\bar{\omega}$ tends to infinity from (\ref{linkfreq}), which is necessary for consistency. Carrying out simulations in order to determine $\varepsilon_c$ for values of $\omega$ in the range $100\leq\omega\leq1000$, we found that $\omega \varepsilon_c = 0.209 \pm 0.004$. 
This is the correct scaling as predicted by (\ref{gamma}) but the constant is not that predicted, $0.3328$. The resolution of this paradox is that in this limit with $\omega=O(\varepsilon^{-1})$, we have $\bar{\omega}  = O(\varepsilon^{-1/2})\to\infty$ from (\ref{linkfreq}). This means that from (\ref{hseries}) $\bar{h}=O(\bar{\omega}^{-1}) = O(\varepsilon^{1/2})$ and so the layer width (\ref{layerwidth}) is of order unity and does not go to zero. The increasing frequency $\bar{\omega}$ is tending to increase the width at the same time as the decreasing $\varepsilon$ is tending to localise the mode, and the two effects cancel out completely. The theory gives the correct scaling law but is not asymptotically correct as the magnetic field is not localised. This indicates the care that has to be taken with double limits and the usefulness of the condition that (\ref{layerwidth}) be small. 
(There may be an asymptotic theory appropriate to the limit $\varepsilon\to0$ with $\varepsilon\omega=O(1)$, based on a reduced, finite number of modes in time $\bar{\tau}$ similar to those in (\ref{fseries})--(\ref{hseries}), but retaining full radial dependence; we leave this for further investigation.)



%

\subsection{Periodic flow with non-zero mean}
\begin{figure}
    \includegraphics[width=1\textwidth]{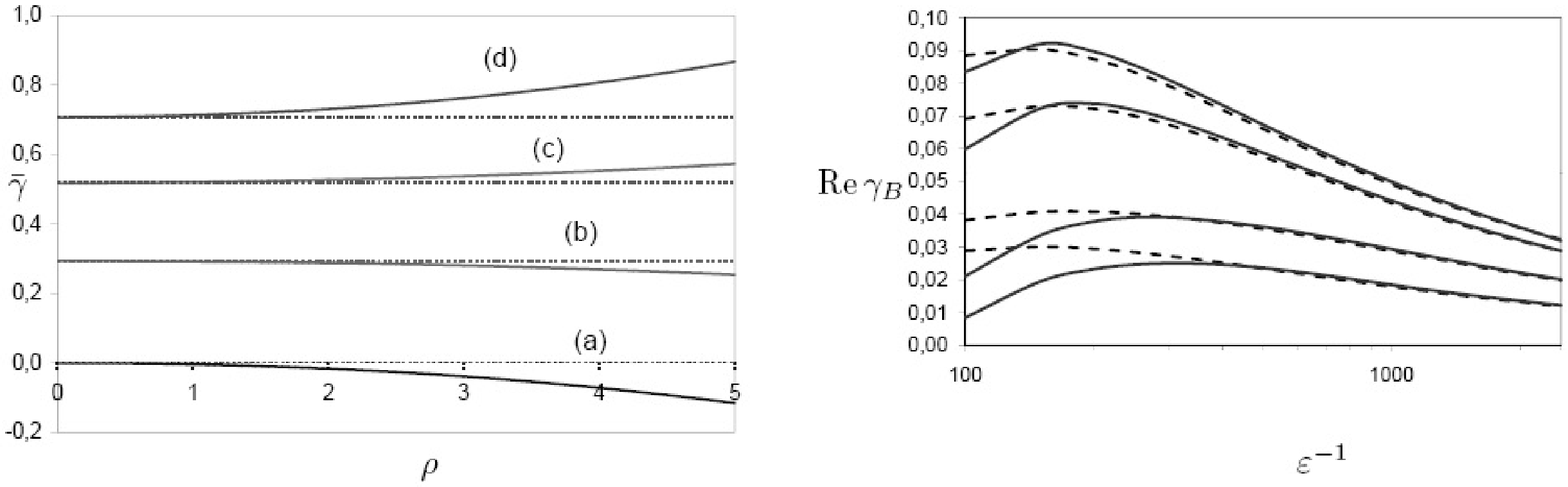}
\caption{Non-zero mean case: growth rate $\bar{\gamma}$ versus $\rho$ (left), for $\bar{\omega}=10$. The curves (a), (b), (c), (d) correspond to  $\mathscr{D}=1, 2 ,3, 4$ respectively.
Growth rate $\Real\gamma_B$ versus $\varepsilon^{-1}$ (right) for $m=1$, $\Gamma = 2$, $k=-0.5$, $ r_0=0.5$, $\omega=100$ and from bottom to top $\rho=1, 4, 8, 10$. The asymptotic results correspond to dashed curves and simulations to full curves.}
\label{fig:NZM}
\end{figure}

Here we consider the case with non-zero mean flow, NZM, in (\ref{eqFforms}) and (\ref{eqforcrescale}). The growth rate $\bar{\gamma}$ from the asymptotic ODEs (\ref{f3}) is plotted against $\rho$ in figure \ref{fig:NZM} (left) for $\bar{\omega}=10$ and different values of $\mathscr{D}$. Taking $\rho=0$ corresponds to the stationary case. Then increasing the fluctuation level $\rho$ may increase the growth rate depending on whether $\mathscr{D}$ is sufficiently large, the transition being for $\mathscr{D}$ between 2 and 3. This shows that fluctuations may increase the dynamo efficiency (at least in the scalings we are using). A different conclusion has been obtained at the dynamo threshold which generally increases with the fluctuation rate \cite{Peyrot07}.
When $\rho$ is increased to large values, the mean part of the flow becomes small compared to the fluctuations. We confirmed that this limit recovers the results of the previous zero-mean case with appropriate rescaling for $\bar{\gamma}$ and $\bar{\omega}$.

In figure \ref{fig:NZM} (right) the growth rate $\Real\gamma_B$ is plotted against $\varepsilon^{-1}$ for different values of $\omega$. The curves show a good agreement between the growth rates from the asymptotic ODEs and from simulation of the full system provided $\varepsilon^{-1}$ is sufficiently large. The difference is less than $4 \%$ for $\varepsilon^{-1} \ge 400$.

\section{Analysis for time-varying resonant radius}\label{secnonres}
\label{sec4}

We now briefly indicate how theory is extended to the more general time dependence, 
\begin{equation}
 \textbf{U}(r,t)=(0,r\Omega(r)[F(t)+qG(t)],V(r) [F(t)-qG(t)]) . 
 \label{velocity1} 
 \end{equation}
Here $q$ is a parameter that controls the difference in time-dependence between the axial and azimuthal components and $F(t)$, $G(t)$ are functions of time, of order unity. For example we could take a general, single frequency, zero-mean case, 
\begin{equation}
F(t) = \cos\omega t , \qquad
G(t) = \cos (\omega t-\Phi) , 
\label{eqFGforms}
\end{equation}

Now generally the resonant radius becomes a function of time, $r_0 (t)$, with 
\begin{equation}
	m\Omega'(r_0) (F(t)+qG(t))+ k V'(r_0) (F(t)-qG(t)) = 0 , 
	\label{resonnance1}
\end{equation}
but such variation is found to have a strong damping effect on the field \cite{Peyrot07}. At the resonant radius the shear in the flow is aligned with the helical field lines in the magnetic mode: in the stationary case, as one departs from this radius, the shear changes direction, tending to introduce fine radial scales and enhanced diffusion. Moving the resonant radius with a time-dependent flow, then, leads to a damping effect on modes, which are strongly suppressed when the field concentration is distant from $r_0(t)$ (noting that the field cannot readily diffuse in radius to follow $r_0(t)$).

For these reasons, in our asymptotic framework we will take $q$ to tend to zero in magnitude as $\varepsilon\to0$. This makes $r_0$ fixed at leading order and we define $r_0$ by equation (\ref{resonnance}) as we did originally. Going through the previous calculations we obtain, in place of (\ref{3r}) and (\ref{3phi}), the equations
\begin{eqnarray}
\lbrack \Dto + c_{0} + ic_1 s \hat{G}(\tau) + i c_{2} s^{2} \hat{F}(\tau) -  \partial_s^{2}  \rbrack    \brho &=& {-2iM}{r_0^{-2}} \bphiho , 
  \label{3r1} \\
\lbrack \Dto + c_{0} + ic_1 s \hat{G}(\tau) + i c_{2} s^{2} \hat{F}(\tau) -  \partial_s^{2}  \rbrack      \bphiho &=&\hat{F}(\tau) r_0 \Omega'(r_0) \brho , 
  \label{3phi1} 
\end{eqnarray}
with the new, linear term defined by 
\begin{equation}
G(t) = \hat{G}(\tau), \quad
q = \varepsilon^{1/3} Q, \quad
c_1 = Q(M\Omega'(r_0) - KV'(r_0)) . 
\end{equation}
Substituting
\begin{equation}
	\left( \begin{array}{l} \brho(\tau,s)\\ \bphiho (\tau,s) \end{array} \right)  
	\exp( c_0 {\tau} )= 
	\left(  \begin{array}{l} a_r \bar{f}(\bar{\tau}) \\  
                                              a_\theta \bar{g}(\bar{\tau}) 
                    \end{array} \right)
	\exp( - a_j \bar{j}(\bar{\tau}) s -  a_h \bar{h}(\bar{\tau})s^2) , 
	\label{layer1}
\end{equation}
and setting $\hat{G}(\tau) =\bar{G}(\bar{\tau})$, and $a_j =c_2^{1/4}$ gives the system,
\begin{eqnarray}
\Dtob \bar{h}  +  4  \bar{h}^{2} &=& i   \bar{F}(\bar{\tau})  , \nonumber \\
\Dtob \bar{j}  +  4\bar{h}\bar{j} &=& i  \mathscr{Q}\bar{G}(\bar{\tau})  , \nonumber \\
\Dtob  \bar{f} -\bar{j}^2 \bar{f}+  2  \bar{h}  \bar{f} &=& -  i  \bar{g} , \label{f3a} \\
\Dtob  \bar{g} -\bar{j}^2\bar{g}+  2  \bar{h}  \bar{g} &=& -  \mathscr{D}  \bar{F}(\bar{\tau}) \bar{f} . \nonumber
\end{eqnarray}
We now have a new parameter that quantifies the difference in the time-dependence of azimuthal and axial flows, and so the variation in resonant radius, given by 
\begin{equation}
 \mathscr{Q} =  c_1c_2^{-3/4} \equiv
 q \varepsilon^{-1/4}  \, \frac{m\Omega'(r_0)-kV'(r_0)}
 {(\tfrac{1}{2}(m\Omega''(r_0) + k V''(r_0)))^{3/4}} \,  . 
 \end{equation}

As for system (\ref{f3}), the system (\ref{f3a}) applies to any time-dependence $\bar{F}(\bar{\tau})$ and $\bar{G}(\bar{\tau})$.
Again this system can be solved numerically to obtain a growth rate. With the specific time-dependence (\ref{eqFGforms}), the growth rate will be a function $\bar{\gamma}( \mathscr{D}, \bar{\omega}, \mathscr{Q}, \Phi)$: the phase angle $\Phi$ quantifies the polarisation of the axial and azimuthal components of the flow, in a loose sense.
 
In the system (\ref{f3a}), we see that changing $\mathscr{Q}$ to $- \mathscr{Q}$ only changes $\bar{j}$ to $- \bar{j}$ without affecting the other variables. Therefore it is sufficient to consider positive values of $\mathscr{Q}$. 
Our numerical investigations, which we summarize rather than presenting graphically, indicate that compared to the curves given in figure \ref{fig:ZeromeanODE} for $\mathscr{Q}=0$,
changing $\mathscr{Q}$ and $\Phi$ systematically leads to lower values of $\bar{\gamma}$, without changing much the shape of the $\bar{\omega}$ and $\mathscr{D}$ dependencies.   
We can show that $\bar{\gamma}$ is $\pi$-periodic in $\Phi$. We find that 
$\bar{\gamma}$ is a maximum for $\Phi=0$, a minimum for  $\Phi=\pi/2$ and that $\bar{\gamma}(\Phi=\pi/4)=\bar{\gamma}(\Phi=3\pi/4)$.
Finally $\bar{\gamma}$ is found to be monotonically decreasing with $\mathscr{Q}$ and it would be interesting to obtain a proof confirming this observation. 

We can investigate these results further by taking the additional limit of large $\bar{\omega}$ as in section \ref{sec3}B. We find that equation (\ref{growthhighfreq}) holds even for $\mathscr{Q}$ non-zero. In other words the leading order growth rate (proportional to $\bar{\omega}^{-1}$) is independent of $\mathscr{Q}$ and $\Phi$ in the limit $\bar{\omega}\rightarrow \infty$ and is given in (\ref{growthhighfreq}). This is very clear in the expansion (\ref{f3a}): the $\bar{j}^2 \bar{f}$ and $\bar{j}^2\bar{g}$ terms come in at one order below what is needed to obtain (\ref{growthhighfreq}). They are asymptotically smaller than $\bar{h}  \bar{f}$ and $\bar{h}  \bar{g}$ in the same equations, as $\bar{h}$ and $\bar{j}$ are both of size $\zeta$ at leading order.

\section{Conclusions}

We extended the theory of the Ponomarenko dynamo in the asymptotic limit of large $\Rm$, to the case of a non-stationary flow. We considered only a very simple model for fluctuating flow, but one that is nevertheless revealing. Within this class of flows it highlights the basic mechanisms for dynamo action, the effect of the motion of the resonant radius in suppressing field generation, and the parameter combinations that are relevant at large $\Rm$. Our results include criteria for dynamo action at large $\Rm$ involving the purely geometrical quantity $\mathscr{D}$, linked to the rate of change of pitch of the velocity shear. For example we find from figure \ref{fig:ZeromeanODE} that the geometrical condition $|\mathscr{D}|>2$ at a given radius is needed for dynamo action with a mode localized there, at high frequencies in the zero-mean case for large $\Rm$.

Note that for stationary flow the corresponding criterion is $|\mathscr{D}|>1$ for magnetic field amplification. To see the relevance of these results, consider the spiral Couette flow, which is simply the general solution for differentially rotating flow forced by rotating, translating cylindrical boundaries 
\begin{equation}
\Omega(r) = A_1 + A_2 r^{-2}, \qquad
V(r) = A_3 + A_4 \log r . 
\end{equation}
This corresponds to $|\mathscr{D}| = 2$, and so satisfies the condition in the stationary case: dynamo action was observed in \cite{Solovyev85}. If the motion of the boundaries is now periodic with zero mean, and sufficiently slow that the above functions are just multiplied by $F(t)=\cos\omega t$, then figure 2 shows that dynamo action becomes marginal provided $\bar\omega$ is large. Of course the full picture for any boundary forcing is complicated by the development of Stokes' layers unless it is slow compared with viscous time-scales. Nonetheless the key point is that flows with larger values of $|\mathscr{D}|$ over a range of radii are likely to be more efficient as dynamos in non-stationary as well as stationary flows, and this consideration could be important in optimizing experiments and understanding experimental or numerical results. We also considered flow fluctuations in which components are out of phase, leading to a time-dependent resonant radius. The results show that this inhibits the dynamo action, confirming previous results obtained for a cellular type of flow \cite{Petrelis06}. 

The fluctuations considered in this paper have a very simple structure, and it would be a natural extension to consider fluctuating flows that carried field across streamlines, in other words depending also on $\theta$ and $z$. This would lose the separation of variables employed here and make the study more numerical, unless averaging is done analytically, which would generally give an alpha effect \cite{Soward}. Other  possible interesting developments would be to consider random time-dependence, which increases the complexity of the system, especially close to the threshold \cite{Leprovost05}, and to include some of the effects of nonlinear feedback on the flow field \cite{Dobler02}. 

\section*{Acknowledgements}

We are grateful to European Network on Electromagnetic Processing of Materials (COST Action P17) which supported a visit of MP to Exeter in 2007, where this research project commenced. AG is grateful for a Leverhulme Trust Research Fellowship held during the completion of this paper. FP is grateful to the Dynamo Program at KITP (supported in part by the National Science Foundation under Grant No. PHY05-51164) for completion of the paper. We thank Prof.\ Andrew Soward and Dr.\ Matthew Turner for helpful comments during this research.

\end{document}